\documentclass[journal,12pt,onecolumn,draftcls]{IEEEtran}

\usepackage{graphicx}
\date{December 11, 2019}

\begin{document}

	\title{Business Model Canvas for Micro Operators in 5G Coopetitive Ecosystem}
	\author{
\IEEEauthorblockN{{ Javane Rostampoor}\IEEEauthorrefmark{1}}, \IEEEauthorblockN{{ Roghayeh Joda}\IEEEauthorrefmark{2}}, and
\IEEEauthorblockN{{ Mohammad Dindoost}\IEEEauthorrefmark{2}\\ }
\and
\IEEEauthorblockA{\IEEEauthorrefmark{1}\small Department of Electrical and Computer Engineering, University of Toronto, Canada},\\  javane.rostampoor@mail.utoronto.ca\\
\IEEEauthorblockA{\IEEEauthorrefmark{2}Communications Technology (CT) Research Faculty, ICT Research Institute, Tehran, Iran},\\ \{r.joda, mdindoost\}@itrc.ac.ir
}
	\maketitle
	
	\begin{abstract}
		In order to address the need for more capacity and coverage in the $5^{th}$ generation (5G) of wireless networks, ultra-dense wireless networks are introduced which mainly consist of indoor small cells. This new architecture has paved the way for the advent of a new concept called Micro Operator. A micro operator is an entity that provides connections and local 5G services to the customers and relies on local frequency resources. We discuss business models of micro operators in a 5G coopetitive environment and develop a framework to indicate the business model canvas (BMC) of this new concept. Providing BMC for new businesses is a strategic approach to offer value to customers. In this research study, BMC and its elements are introduced and explained for 5G micro operators.
	\end{abstract}

	\section{Introduction}
	\subsection{Motivation}
	Wireless communication networks have been facing a tremendous growth in data traffic and the number of customers seeking for connections and services \cite{5G}. Cisco Visual Networking Index (VNI) has anticipated that by 2021 when the first pioneer operators are deploying 5G, data traffic will increase roughly seven times as it was in 2016 \cite{cisco}. The significant fact is that Cisco forecasts claim that $\% 96$ of the traffic will be consumed indoors. In this regard, the success of the next generation wireless network in dealing with the growing data demand is closely dependent on its approach to handle indoor traffic. 5G solution to tackle this challenge is densifying the network and deploying small cells, especially in indoor hotspots \cite{dense}.

The increasing need for deploying ultra-dense indoor small cell networks has introduced a new concept in 5G, called micro operator (MO). A micro operator is an entity that provides its customers with connectivity and local services, confined to a local narrow operation area and dependent on local spectrum resources \cite{micro}. On the other hand, as MO will bring new opportunities and challenges in 5G networks, it should be studied from a business perspective. Business model canvas (BMC) is a suitable framework for new businesses and companies to manage their resources and strategies and analyze their business activities \cite{canvas}.

As MOs rely on other available resources such as site and spectrum, they have to operate in a cooperative manner with other MOs or other partners. In addition, as they have to compete with each other to serve customers and make revenues, they have to operate in a cooperation-competition (coopetition) environment.
Coopetition is a relationship between multiple business actors that are simultaneously involved in cooperation and competition relationships \cite{micro, coopetition}.
On the other hand, understanding a typology to clarify and classify different business models for newly introduced technologies is a critical factor that makes analysis much easier. The 4C model is considered a typology that can cover a majority of activities on the Internet. 4C stands for connection, content, context, and commerce which are four different business models of Internet-based technologies and are closely related to each other. The ecosystemic nature of many Internet-based technologies is reflected in 4C layered model and each layer has an influence on the other layers \cite{4c, 4c2}.

This paper aims to provide BMC for MOs based on their ecosystemic nature. Hence, the new technology can be more justifiable and more practical to play a vital role in 5G networks. Note that as BMC covers the 4C model as a part of it, BMC is much more precise than just using 4C to explain MOs.
	
	\subsection{Related Works}
	This work is founded on the intersection of two research areas: 1) MOs and their coopetitive role in 5G networks, and 2) BMC as a strategic template for developing technologies.
The first notion of micro operators is introduced in \cite{micro}, where the authors define a micro operator as an entity that: 1) provides its customers with local connectivity and local services, 2) is confined to a limited and specific operation area and 3) is highly dependant on the limited spectrum resources. The paper discusses coopetitive business models of micro operators based on 4C model.
The authors in \cite{micro2} investigate the concept of micro operators and discuss its regulatory issues and technical considerations to find out the feasibility of this concept in practice. In this regard, the authors study the business models of micro operators in terms of providing customers with connectivity, content, context, and commerce.
Micro operators are introduced in \cite{micro3} from a more technical point of view and cloud-based deployment to enable connection and service provisioning is discussed. In addition, some challenges such as security and regulatory issues are studied.

Spectrum as a critical required resource for the deployment of micro operators is studied in \cite{ spectrum3, spectrum1, spectrum2}.
Micro licensing is defined in \cite{spectrum3} which means granting local access rights as licenses to deploy 5G networks in limited operating areas. The authors provide micro-licensing elements that should be considered by regulators to deploy 5G networks.
 Micro operators presented in \cite{spectrum1}, can buy multiple subbands from the regulator and each small cell deployed by micro operators, can utilize multiple licensed subbands. In order to satisfy the quality of service (QoS) of users, the authors introduce a greedy algorithm to find out the number of licensed subbands and which subband the micro operator should purchase from the regulator.
As micro operators rely on some local licensed spectrum, spectrum sharing is a fundamental part of this new concept. An important issue in spectrum sharing is interference management so that the 5G network remains free from distractive interference. The authors in \cite{spectrum2}  characterize interference scenarios between different micro operators and focus on building-to-building scenarios that licensees in two separate buildings in co-channel and adjacent channel cases. This is a probable scenario for micro operators since it allows a myriad of stakeholders to gain access to the spectrum and play the role of micro operators in 5G networks.
The interference characteristics are studied and the allowable transmit power of each base station is analyzed as a function of the minimum separation distance between micro operators \cite{spectrum2}.

As it is mentioned, the business model canvas is a strategic approach for businesses to provide their customers with value.
Business model canvas for IoT systems is investigated in \cite{IoT}, in which the authors undertook case studies of some IoT companies to validate the business model.
In \cite{new}, three generic 5G micro operators business models: Vertical, Horizontal, and Oblique are discussed and they are compared with each other from scalability and adaptability perspectives. In fact, the authors focused on the differences between micro operators' business models and general mobile network operators' business models.

	\subsection{Contribution}
	In this paper, the concept of micro operators in an ecosystemic coopetitive environment is investigated in future 5G networks and a framework for its business model is presented. This framework is a business model canvas which is a principal part of all new businesses to visualize their future results based on the value flows from the owners of the business to the customers. As discussed in related works, although there exist some studies that cover the micro operator's concept and its ecosystemic nature, the literature lacks a solid framework in terms of the business model canvas of this concept. In this regard, we propose a business model canvas for 5G micro operators which has not been already studied in the literature.
	\subsection{Paper Structure}
	The rest of this paper is organized as follows. Section II represents the business model canvas template and each component of it. Business model canvas for 5G micro operators is investigated in Section III and the paper is concluded in Section IV in which, recommended future work is explained.
	\section{Business Model Canvas}
Business model canvas is an important tool to analyze and visualize business models. The canvas represents how a company can provide its customers with value and how revenue streams flow to the company. This model is a helpful strategy to describe, analyze, and understand a business, and hence sketching a business model canvas is highly recommended for new starting businesses. It consists of nine blocks: key partners, key activities, key resources, value proposition, customer relationships, channels, customer segments, cost structure, and revenue streams \cite{canvas1, canvas2}. These nine blocks can be interpreted as four major blocks, i.e. infrastructure, product, interface to customers, and financial aspects \cite{canvas3}. The infrastructure block explains \emph{how} a value is created for customers and it contains `key activities', `key partners', and `key resources' blocks. The product block represents \emph{what} value is created and it includes the `value proposition' block. The interface to customer block explains the interface \emph{for whom} the value is created. It contains three blocks: `customer relationships', `channels', and `customer segments'. The last block deals with \emph{financial issues} of value creation and distribution and comprises `cost structure' and `revenue streams'.
 Fig. 1 represents a business model canvas and its blocks.

	\begin{figure}[!t]
		\centering
		\includegraphics[trim= 0mm 117mm 0mm 0mm,clip, width=3.4in]{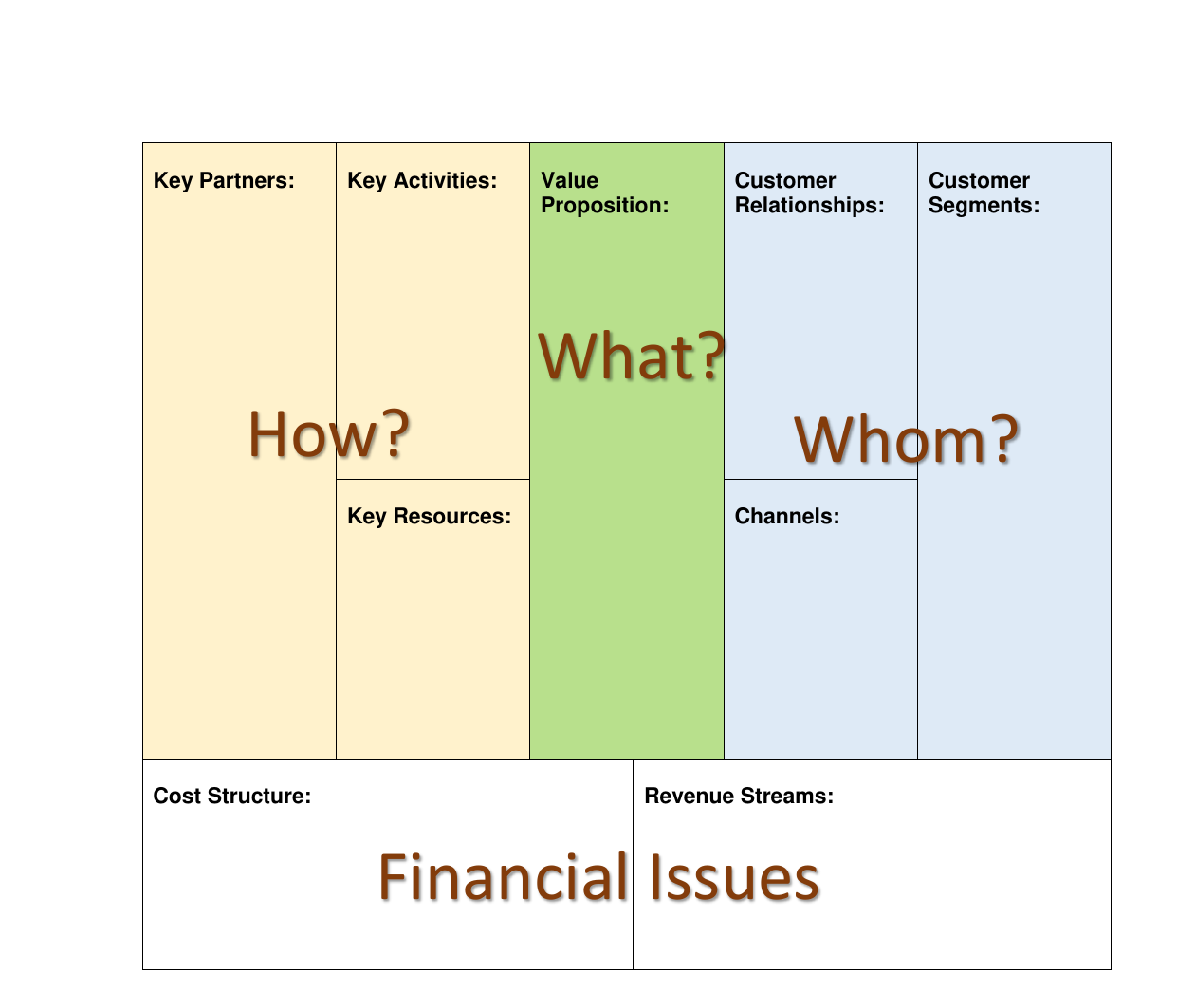}
		\caption{Business Model Canvas.}
		\label{fig1}
	\end{figure}

	\section{Business Model Canvas for Micro Operators}
	In this section, we propose a framework to discuss the business perspective of micro operators as a new player in the 5G environment. As explained in the previous section, the business model canvas is adopted from \cite{canvas2} and it consists of nine blocks which should be filled in for micro operators. Each block is introduced and mapped to the micro operator business model.
\subsection{Key Partners}
The first block is key partners and describes the network of suppliers and stakeholders who take part in the business. The authors in \cite{micro2}  consider six main stakeholders for micro operators. However, they believe that as micro operator is a new concept in the business environment, it is open to new stakeholders to take part in and play a role. The main stakeholders are defined as follows.

Facility users and owners are the most important stakeholders who give the micro operator permission for the deployment of small cells in the site. Regulatory organizations perform spectrum auctions and issue micro licenses for the micro operators. In addition, they define regulations under which micro operators should act such as the level of sharing between operators, the operation area, the maximum transmit power of base stations, and the maximum level of interference, etc. In fact, regulators have a critical role and have to receive applications from micro operators and respond to them within a specific period of time. Additionally, the regulatory is responsible for granting access rights to the spectrum among the 5G applicants in a non-discriminatory manner. For micro operators, this process occurs in micro licensing which includes offering local temperate access rights to 5G spectrum to deploy small cell networks with limited coverage. In micro licensing, license holders are micro operators, and access rights of specific parts of the spectrum are granted to them with some level of guaranteed interference protection. Micro license duration is shorter than exclusive access rights and can be extended depending on the situation. Different from the traditional exclusive licenses, micro-licensing buildings play a critical role and influence on the area of operation. In fact, based on indoor and outdoor operation areas and interference regimes, the licenses are granted. As licenses are granted in limited operation areas, the price of licenses is lower than exclusive spectrum access rights and is defined in order to increase competition between operators in fact, maximizing regulatory revenues is not the aim. In addition, regulations have to define rules for transferring access rights to new entrants and their considerations. Regulatory should explain under which conditions a micro operator can transfer its rights to a third party and the possibility of applying the access right in a new location (portability). Unlike traditional licensing that has nationwide coverage obligations, micro licensing imposes coverage obligations within a defined local coverage area \cite{spectrum3}.

In addition, as micro licenses are granted locally, there might exist some incumbent users of spectrum and license holders. In this situation, micro operator is responsible for protecting the incumbents from harmful interference. On the other hand, interference coordination among micro license holders is another substantial issue that should be considered and managed by micro operators. All these interference coordination should be monitored by regulatory. Hence, regulatory is one of the most important stakeholders in 5G micro operators.

 Network infrastructure vendors provide micro operators with network infrastructures and are one of the key partners. Device and equipment vendors provide customers with devices capable of connecting to the micro operators and also provide micro operators with the required equipment. In fact, technological enablers of the micro operator concept are vendors who facilitate 5G implementations by removing technological barriers.

 Content providers offer content to micro operators in order to be accessed by customers of them. Mobile network operators (MNOs) are another important stakeholders that have a critical role in the 5G micro operator environment. Micro operators can make use of MNOs' infrastructures through some agreements with them and micro operators can benefit MNOs by offering their customers connectivity (cooperation relationship). However, these operators have concerns about maintaining their customers in case of offering parallel connectivity services to them (competition relationship). Hence, the relationship between MNOs and micro operators lies in the area of coopetition relationships and follows the rules of it.

It should be noted that 5G micro operator stakeholders are not confined to the mentioned list. By implementing micro operators in a real environment, some other beneficiaries and third parties may play a role in this environment. In fact, micro operator concept is open to newcomers to take part in the business and serve customers.
\subsection{Key Activities}

This block represents the most important activities that a business owner must do to offer value to its customers. According to \cite{micro}, 4C business model typology is applicable to  5G micro operators due to their ecosystemic nature them. In this regard, the activities associated with these four levels can be considered the key activities of the micro operator in the business model canvas. Accordingly, the micro operator's key activities are to provide customers with connection, content, context, and commerce.

\subsection{Key Resources}
This block describes the most important assets and resources required by the business model. These assets include infrastructures that can assist network establishment and connectivity, sites for installing small cells, licenses achieved by regulators, and contents created by users, content providers, or micro operator. In addition, intellectual property is another asset that is needed for every developing technology to enable innovation and new features. A fast and reliable internet connection is another asset that is required for ensuring the robustness of backhaul connection as well as its capability to transfer data traffic as speedy as possible.

\subsection{Value Proposition}
The value proposition block describes the key value of products or services that are conveyed to the customers. Usually, it explains the reason for introducing this new business and why the new products or services can stand out among previous peers. The micro operator concept stems from the fact that small cells are an integral part of future 5G networks while in some situations, MNOs are reluctant to build new small cell sites. In addition, micro operators are introduced to break the monopoly on small cell services to open new opportunities and enhance customer experience. In fact, as micro operators increase the number of stakeholders who can participate in the whole business, they are capable of reducing monopoly by increasing competition among stakeholders. Accordingly, micro operators can benefit customers in the way small cells can, such as by providing them with more reliable capacity and coverage, reduced latency, accelerating development, and making feasible new technologies such as slicing, virtualization, etc. \cite{smallcell}. Slicing and virtualization lead to personalization and satisfy their special needs,  expected QoS, and quality of experience (QoE).

\subsection{Customer Relationships}
This block represents the type of relationships between a company and its customers and it leaves a significant influence on the customers' experience. Co-creation is a strategy that can be developed through active collaboration of stakeholders. It is one of the most important strategies in business to bring different parties together and share ideas to create innovation and address the needs \cite{cocreation}. In this regard, Co-creation is a beneficial strategy for micro-operators to get feedback from the customers and receive the customers' and reviewers' viewpoints to be able to expand their business. Co-creation in this business model makes value creation and personalization feasible for different groups of customers with each group's special needs.
	\begin{figure*}[!t]
		\centering
		\includegraphics[trim= 0mm 77mm 0mm 0mm,clip, width=\textwidth]{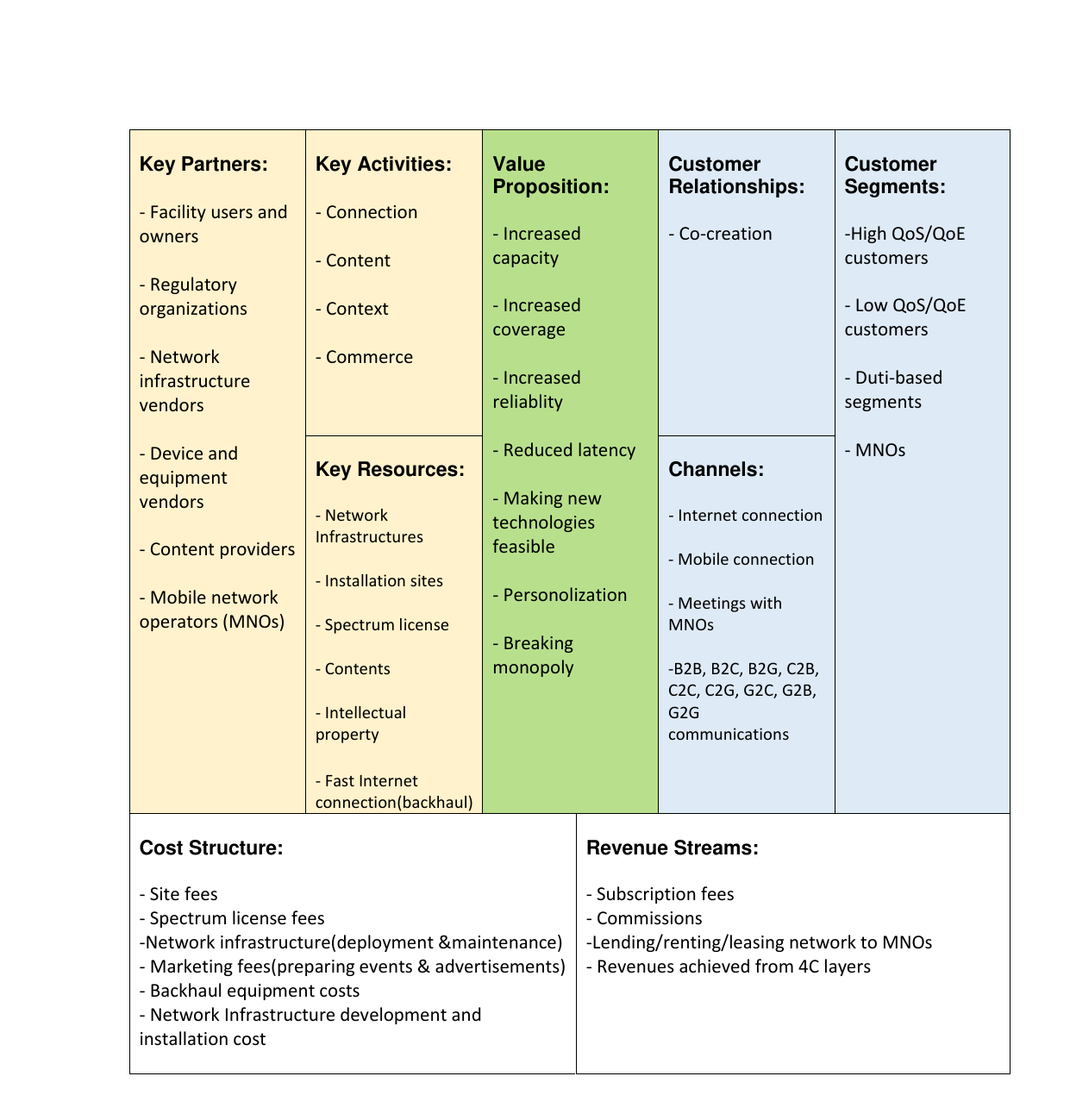}
		\caption{Business Model Canvas for 5G Micro Operators.}
		\label{fig1}
	\end{figure*}
\subsection{Channels}
This block explains how a company reaches its customers and what are the interfaces between them for value transfer. Value can be transferred by Internet through mobile or fixed communications to the customers in the majority of scenarios. In cases where micro operators plan to serve MNO's customers, value can be communicated by meetings between MNOs and micro operators to reach an agreement. When micro operators are responsible for providing customers with commerce, channels can be business-to-business (B2B) communications, business-to-consumer (B2C), business-to-government (B2G), as well as consumer-to-business (C2B), consumer-to-consumer (C2C), consumer to government (C2G), government to business (G2B),
government to consumer (G2C) and government to government (G2G) communications. Different types of e-commerce communications are investigated in \cite{b2b,b2g}. In fact, as micro operators can provide connections, any e-commerce between customers, businesses, and governments can be considered as communication assisted by them.
\subsection{Customer Segments}
This block describes different groups of customers micro operator aims to reach and transfer value to them. The segmentation can be based on common needs, common location, or common position. Micro operators can categorize their customers based on their required QoS and QoE. As micro operators can reach agreements with MNOs to serve their customers, MNOs can be considered as micro operators' customers. In this regard, one important segment of this block is dedicated to MNOs.

It should be noted that customer segmentation is not necessarily based on connection level (QoS and QoE). Micro operators can categorize their customers based on other needed values, likewise. For instance, a micro operator who has deployed a small cell in a university can categorize its customers into groups based on their university status such as students, professors, university personnel, etc. In this case, the micro operator can manage and deliver values to different segments efficiently. In fact, these segments require different content and one of the responsibilities of micro operators is to provide their customers with content.

\subsection{Cost Structure}
Cost structure describes the most important costs required to operate a business. As micro operators have to get permission from site owners to access the site and establish their own business, one of the most important costs is site fees (in case the micro operator is not the site owner). Another cost is acquiring the right to access spectrum which is paid to the regulatory organizations via licence fee. Although, micro operators utilize micro licenses to establish their networks which are less expensive, the spectrum fees are important parts of a micro operator's cost structure. Furthermore, micro operators have to pay for network infrastructures and their maintenance. Other costs include operating costs such as data analysis and data storage costs. As micro operators have to reach agreements with MNOs to serve their customers in some special cases, marketing costs would be an important factor in this case, since preparing some events and advertisements to introduce their business is costly. As denoted, another key resource that micro operators need to establish their business is a backhaul link. This link mainly relies on broadband connections of the premises and micro operators have to pay for utilizing the link. In some other cases, backhaul links are based on microwave links which have to be permitted by regulatory in case of operating in licensed spectrum bands.
\subsection{Revenue Streams}
This block represents the main revenue streams that a company receives from its business strategy. Revenue streams can be achieved through one-time transactions or ongoing payments in the future \cite{canvas1}. Micro operators can obtain revenues by agreements with MNOs to serve their customers or by serving their own customers and receiving subscription fees. In the case of reaching an agreement with MNOs, micro operators can earn revenues by renting/lending or leasing their network. According to the 4C business model typology, micro operators can gain revenues from all 4C levels. At the connection level, subscription fees are received from customers. In content and context layers, customers are charged in exchange for receiving their required content or context. In the commerce layer, the micro operator can charge commission on done transactions by their customers.

Fig. 2 summarizes the mentioned blocks and illustrates the business model canvas for micro operators in 5G networks.

	\section{Conclusion and Future Work}
	In this paper micro operators' role in 5G future networks and their potential opportunities were discussed. Micro operator as a new business has to be studied from business model perspectives in order to clarify and explain its prospective role. In this regard, micro operators and their coopetitive nature were studied as well as 4C business model typology. Then, we proposed a business model canvas for 5G micro operators and explained each block, which considers the 4C model as a part of it. Based on the micro operator's ecosystemic nature and business opportunities, each block was filled and justified. Accordingly, we provided a solid framework for a business that is willing to establish a 5G micro operator. As this concept is not deployed yet, the business model canvas is a visualized framework for its implementation. In this regard, in the future by the appearance of micro operators, the study can be updated by some new evidence and interviews with the stakeholders.


\end{document}